\documentclass[12pt]{iopart}

\usepackage{iopams} 
\usepackage{graphicx}
\usepackage{inputenc}
\usepackage{tablefootnote}
\usepackage{cite}

\graphicspath{{}}

\begin{document}
\title[]{
STE-QUEST - Test of the Universality of Free Fall Using Cold Atom Interferometry}
\author{
D.~Aguilera\textsuperscript{1}, 
H.~Ahlers\textsuperscript{2},
B.~Battelier\textsuperscript{3},
A.~Bawamia\textsuperscript{4}, 
A.~Bertoldi\textsuperscript{3},
R.~Bondarescu\textsuperscript{5}, 
K.~Bongs\textsuperscript{6}, 
P.~Bouyer\textsuperscript{3},
C.~Braxmaier\textsuperscript{1,7},
L.~Cacciapuoti\textsuperscript{8},
C.~Chaloner\textsuperscript{9},
M.~Chwalla\textsuperscript{10},
W.~Ertmer\textsuperscript{2},
M.~Franz\textsuperscript{11},
N.~Gaaloul\textsuperscript{2},
M.~Gehler\textsuperscript{8},
D.~Gerardi\textsuperscript{10},
L.~Gesa\textsuperscript{12},
N.~G\"urlebeck\textsuperscript{7},
J.~Hartwig\textsuperscript{2}, 
M.~Hauth\textsuperscript{13},
O.~Hellmig\textsuperscript{14},
W.~Herr\textsuperscript{2}, 
S.~Herrmann\textsuperscript{7},
A.~Heske\textsuperscript{8},
A.~Hinton\textsuperscript{6},
P.~Ireland\textsuperscript{9},
P.~Jetzer\textsuperscript{5}, 
U.~Johann\textsuperscript{10},
M.~Krutzik\textsuperscript{13},
A.~Kubelka\textsuperscript{7}
C.~L\"ammerzahl\textsuperscript{7}, 
A.~Landragin\textsuperscript{15},
I.~Lloro\textsuperscript{12},
D.~Massonnet\textsuperscript{16},
I.~Mateos\textsuperscript{12},
A.~Milke\textsuperscript{7},
M.~Nofrarias\textsuperscript{12},
M.~Oswald\textsuperscript{11},
A.~Peters\textsuperscript{13},
K.~Posso-Trujillo\textsuperscript{2},
E.~Rasel\textsuperscript{2},
E.~Rocco\textsuperscript{6},
A.~Roura\textsuperscript{17},
J.~Rudolph\textsuperscript{2},
W.~Schleich\textsuperscript{17},
C.~Schubert\textsuperscript{2},
T.~Schuldt\textsuperscript{1,11},  
S.~Seidel\textsuperscript{2},
K.~Sengstock\textsuperscript{14},
C.~F.~Sopuerta\textsuperscript{12},
F.~Sorrentino\textsuperscript{18}, 
D.~Summers\textsuperscript{9},
G.~M.~Tino\textsuperscript{18}, 
C.~Trenkel\textsuperscript{19},
N.~Uzunoglu\textsuperscript{20}, 
W.~von~Klitzing\textsuperscript{21},
R.~Walser\textsuperscript{22}, 
T.~Wendrich\textsuperscript{2},
A.~Wenzlawski\textsuperscript{14},
P.~We{\ss}els\textsuperscript{23}, 
A.~Wicht\textsuperscript{4},
E.~Wille\textsuperscript{8}, 
M.~Williams\textsuperscript{19}, 
P.~Windpassinger\textsuperscript{14}, 
N.~Zahzam\textsuperscript{24}
}

\address{
\textsuperscript{1} German Aerospace Center (DLR), Institute for Space Systems, Robert-Hooke-Str.~7,
28359 Bremen, Germany\\
\textsuperscript{2} Institute of Quantum Optics, Leibniz University Hanover,
Welfengarten 1, 30167 Hanover, Germany\\
\textsuperscript{3} Laboratoire Photonique, Num\'erique et Nanosciences-LP2N Universit\'e Bordeaux-IOGS-CNRS: UMR 5298, Talence, France\\
\textsuperscript{4} Ferdinand-Braun-Institut, Gustav-Kirchhoff-Str. 4, 12489 Berlin, Germany\\
\textsuperscript{5} Institute of Theoretical Physics, University of Zurich,
Winterthurerstr. 190, 8057 Zurich, Switzerland\\
\textsuperscript{6} School of Physics and Astronomy, University of Birmingham, Birmingham, B15 2TT, United Kingdom\\
\textsuperscript{7} Center of Applied Space Technology and Microgravity
(ZARM), University Bremen, Am Fallturm, 28359 Bremen, Germany\\
\textsuperscript{8} ESA - European Space Agency, ESTEC, Keplerlaan 1, 2200 AG
Noordwijk ZH, Netherlands\\
\textsuperscript{9} SEA House, Bristol Business Park, Coldharbour Lane, Bristol BS16 1EJ, United Kingdom\\
\textsuperscript{10} Astrium GmbH - Satellites, Claude-Dornier-Str., 88090 Immenstaad, Germany\\
\textsuperscript{11} Institute of Optical Systems, University of Applied
Sciences Konstanz (HTWG), Brauneggerstr. 55, 78462 Konstanz, Germany\\
\textsuperscript{12} 
Institut de Ci\`encies de l'Espai (CSIC-IEEC), Campus UAB, Facultat de
Ci\`encies, 08193 Bellaterra, Spain\\
\textsuperscript{13} Humboldt-University Berlin, Institute for Physics,
Newtonstr. 15, 12489 Berlin, Germany\\
\textsuperscript{14} Institute of Laser-Physics, University of Hamburg, Luruper
Chaussee 149, 22761 Hamburg, Germany\\
\textsuperscript{15} SYRTE- Observatoire de Paris, 61 avenue de l'obsevatoire,
75014 Paris, France\\
\textsuperscript{16} CNES - Centre national d'etudes spatiales, 2 place Maurice
Quentin, 75039 PARIS CEDEX 01, France\\
\textsuperscript{17} Department of Quantum
Physics, University of Ulm, Albert-Einstein-Allee 11, 89081 Ulm, Germany\\
\textsuperscript{18} Dipartimento di Fisica e Astronomia and LENS Laboratory,
Universit{\`a} di Firenze - INFN, Sezione di Firenze - via G. Sansone 1, 50019
Sesto Fiorentino (Firenze), Italy\\
\textsuperscript{19} Astrium Ltd, Gunnels Wood Road, Stevenage SGI 2AS, United
Kingdom\\
\textsuperscript{20} National Technical University of Athens, 28 Oktovrio 42,
10682 Athens, Greece\\
\textsuperscript{21} Institute of Electronic Structure and Laser, Foundation for
Research and Technology - Hellas, P.O. Box 1527, 6R-71110 Heraklion, Greece\\
\textsuperscript{22} Institut f\"ur Angewandte Physik, Technische Universit\"at
 Darmstadt, Hochschulstr. 4a, 64289 Darmstadt, Germany\\
\textsuperscript{23} LZH - Laser Zentrum Hannover e.V., Hollerithallee 8, 30419 Hannover, Germany\\
\textsuperscript{24} ONERA - Office National d'Etude et de Recherches
Aerospatiales, Chemin de la Huniere, 91761 Palaiseau, France}

\ead{
gaaloul@iqo.uni-hannover.de, norman.guerlebeck@zarm.uni-bremen.de,
thilo.schuldt@dlr.de }

\begin{abstract}
The theory of general relativity describes macroscopic phenomena driven by the influence of gravity while quantum mechanics brilliantly accounts for microscopic effects. Despite their tremendous individual success, a complete unification of fundamental interactions is missing and remains one of the most challenging and important quests in modern theoretical physics.
The STE-QUEST satellite mission, proposed as a medium-size mission within the
Cosmic Vision program of the European Space Agency (ESA), aims for testing
general relativity with high precision in two experiments by performing a
measurement of the gravitational redshift of the Sun and the Moon by comparing terrestrial
clocks, and by performing a test of the Universality of Free Fall of matter
waves in the gravitational field of Earth comparing the trajectory of two Bose-Einstein
condensates of \textsuperscript{85}Rb and \textsuperscript{87}Rb.
The two ultracold atom clouds are monitored very precisely thanks to
techniques of atom interferometry. This allows to reach down to
an uncertainty in the E\"otv\"os parameter of at least $2\cdot10^{-15}$. In this
paper, we report about the results of the phase A mission study of the atom
interferometer instrument covering the description of the main payload
elements, the atomic source concept, and the systematic error sources.
\end{abstract}

\vspace{2pc}
\submitto{\CQG}

\section{Introduction}
The current theory of gravity, general relativity, is based on
Einstein's Equivalence Principle. It consists of three parts: The Universality of
Free Fall (UFF), the Local Position Invariance, and the Local Lorentz
Invariance. 

The Universality of Free Fall (UFF),\footnote{The UFF is also called Weak
Equivalence Principle.} implies that the trajectories of test masses, for which
tidal deformations, self-gravity, electromagnetic charges, spin, etc. are
negligible, depend only on their initial position and their initial velocity.
The Local Lorentz Invariance postulates that the the outcome of any
non-gravitational experiment performed in a freely falling frame is independent
of the velocity and of the orientation of that frame. The Local Position
Invariance states that the outcome of such an experiment is also independent of
where and when in the universe it is carried out, cf.
\cite{Will_2006,Lammerzahl_2011,Liberati_2013}.

Essentially all efforts to unify gravity with the other fundamental interactions
(e.g.\ string theory, canonical quantum gravity, etc.) predict a violation of
Einstein's Equivalence Principle at some scale 
\cite{Lammerzahl_2011,Kiefer_2012}.
Therefore, a crucial step towards an unified theory requires experiments that
test the assumptions and principles of general relativity and search for
possible violations or set bounds to the possible deviations. Such deviations
and also their absence could, indeed, shed some light on the quantum nature of
gravity. This holds in particular for the low energy limits of string theory,
where extra moduli fields arise, see, e.g.,
\cite{Damour_1994,Damour_2002,Damour_2012}. Moreover, theories with a fifth
force, theories invoked to explain dark energy, and theories with varying
fundamental constants and non-minimal coupling can entail a violation of the
UFF, see, e.g., \cite{Fischbach_1986,Fischbach_1998},
\cite{Brans_1961,Dvali_2002,Puetzfeld_2013}, and
\cite{Carroll_2010,Wetterich_2003,Chen_2005}, respectively. Another violation
scenario is described in \cite{Goeklue2008}. A phenomenological framework describing a violation of the UFF is, for example, the Standard Model Extension, see, e.g., \cite{Will_1993,Colladay_1998}. A violation of the UFF is quantified by the
E\"otv\"os ratio $\eta=|\Delta a|/|\vec{g}\cdot\vec{e}_{\Delta a}|$, where
$\Delta a$ denotes the differential acceleration of the two test bodies and
$\vec{g}\cdot\vec{e}_{\Delta a}$ the projection of the local gravitational
acceleration $\vec{g}$ onto the sensitive axis $\vec{e}_{\Delta a}$ of the accelerometer.
 
STE-QUEST (Spacetime Explorer and Quantum Equivalence Principle Space Test) is a
medium-size (M3) candidate satellite mission, which we proposed to ESA in the
scope of the Cosmic Vision program.
It is currently in the assessment phase (Phase A).
The planned STE-QUEST satellite consists of a dual species
(\textsuperscript{85}Rb and \textsuperscript{87}Rb) atom interferometer (AI) and
a microwave link. A microwave clock based on laser cooled Caesium atoms and an
optical link are considered as an optional payload.
The AI shall test UFF with atomic wave packets of different masses to the unprecedented accuracy of about $\eta \leq 2\cdot 10^{-15}$. The use of Bose-Einstein condensates (BEC) allows taking advantage of their intrinsic properties, i.e. long coherence length and slow expansion. The microwave link will allow for a measurement of the gravitational redshift due to
the Sun's and the Moon's gravitational potential by ground clock comparison,
expected to reach an uncertainty of $5\cdot 10^{-7}$ and $9\cdot 10^{-5}$,
respectively. The former is presently measured to the few percent level
\cite{LoPresto_1991,Krisher_1993}; the latter is not experimentally determined
yet. In case the optional atomic Caesium clock is included in the STE-QUEST
payload, the redshift due to the Earth's gravitational field will also be
measured with an uncertainty of $2\cdot 10^{-7}$ resulting in a factor $350$
improvement over the current best measurements by Gravity Probe A
\cite{Vessot_1980}. The long common-view contacts required to compare ground
clocks and the need for a strong gravity field for maximizing an eventual
UFF-violating signal define the highly elliptic STE-QUEST orbit.
The mission details were investigated in an independent industry study; its
results are presented in \cite{STE_QUEST_mission}. Preliminary aspects of the
mission were presented in a recent conference proceedings \cite{Tino2013}.

\begin{table}[h!tb]
\begin{center}
 \vspace{0.2cm}  
 \begin{tabular}{lclc}
 \hline 
  Apparatus & Target precision for $\eta$ & Species & Ref. \\
  \hline 
  Torsion balance$^{3)}$ & $(0.3\pm1.8)\cdot10^{-13}$ & Ti, Be & \cite{Schlamminger08PRL} \\ 
  Lunar Laser Ranging$^{2,3)}$ & $(-0.8\pm1.8)\cdot10^{-13}$ & Moon, Earth &
  \cite{Williams2012CQG}\\
  AI/FG5 & $(7\pm7)\cdot10^{-9}$ & Cs, Glass & \cite{Peters99Nat}  \\ 
  Dual AI (Garching)& $(1.2\pm1.7)\cdot10^{-7}$ & \textsuperscript{85}Rb, \textsuperscript{87}Rb & \cite{Fray2004PRL} \\
   Dual AI (ONERA)& $(1.2\pm3.2)\cdot10^{-7}$ & \textsuperscript{85}Rb, \textsuperscript{87}Rb & \cite{Bonnin2013PRA} \\
   Dual AI (Firenze)& $7\cdot10^{-7}$ & \textsuperscript{87}Sr, \textsuperscript{88}Sr & \cite{Tino2012Varenna} \\
  Dual AI$^{1)}$ (Hanover) & $10^{-9}$ & \textsuperscript{85}Rb, K & \cite{IQ13website}  \\ 
  Dual AI$^{1)}$ (Berkeley) & $10^{-14}$ & \textsuperscript{6}Li, \textsuperscript{7}Li & \cite{Hamilton2012APS}  \\
  Dual AI tower initial/upgrade$^{1)}$ (Stanford)  & $10^{-15}$/$10^{-16}$ &
  \textsuperscript{85}Rb, \textsuperscript{87}Rb & \cite{Kasevich2007}\\
  \hline
 \end{tabular}
  \caption{Existing and planned UFF tests on ground. $1)$ Work in progress. $2)$
 LLR references the differential acceleration between Moon and Earth to the
 gravitational field of the Sun. All other tests in this table are referenced to
 the gravitational field of Earth. $3)$ Macroscopic
 test masses.\label{tab:TestsOnGround} } 
 \end{center}
 \end{table}

 \begin{table}[h!tb]
 \begin{center}
 \vspace{0.2cm}  
 \begin{tabular}{lclc}
 \hline
 Apparatus & Target precision for $\eta$ & Species & Ref. \\ 
 \hline 
 SAI ground based/in zero-g & $[10^{-7}$/$1.8\cdot10^{-10}]$ $^{2)}$
 & \textsuperscript{87}Rb & \cite{Sorrentino2010MST} \\
 ICE & $10^{-11}$ & \textsuperscript{87}Rb, K & \cite{Bouyer2009} \\ 
 QUANTUS & $6.3\cdot10^{-11}$ & \textsuperscript{87}Rb, K & \cite{Rudolph2011MST} \\ 
 MICROSCOPE$^{1)}$ & $10^{-15}$ & Pt, Ti & \cite{Touboul12CQG} \\ 
  STEP$^{1)}$ & $10^{-18}$ & Pt, Ir, Nb, Be & \cite{Overduin12CQG} 
  \\
  GG$^{1)}$ & $10^{-17}$ & $^{3)}$ & \cite{Nobili12CQG}  \\ 
  \hline
 \end{tabular}
  \caption{\label{tab:TestsZerog}Planned and proposed UFF tests in space and
  zero-g environments. All tests in this table are referenced to the gravitational field
 of Earth. $1)$ Macroscopic test masses. $2)$ Single species experiment, sensitivity given in
 m\,s$^{-2}$ Hz$^{-1/2}$. $3)$ Not yet decided.}
 \end{center}
 \end{table}

Many tests of the UFF on ground and in micro-gravity environments reported no
violation down to the $1\cdot 10^{-13}$ level; we summarized them in Tab. \ref{tab:TestsOnGround} and Tab.
\ref{tab:TestsZerog}.
STE-QUEST performs a \emph{quantum} test of the UFF by tracking the propagation
of matter waves in Earth's gravitational field by means of a two species atom
interferometer achieving an accuracy of at least $2\cdot 10^{-15}$. The matter waves
are generated from two ensembles of Rubidium isotopes
(\textsuperscript{85}Rb and \textsuperscript{87}Rb), which are cooled down until
Bose-Einstein condensation sets in, allowing an improvement of the UFF test by
orders of magnitude compared to the non-condensate matter case, see \cite{Fray2004PRL}.
The interferometer is based on previous studies like SAI (Space Atom
Interferometer) \cite{Sorrentino2010MST}, SpaceBEC (Quantum gases in
microgravity), the french CNES project I.C.E. (Interf\'erom\'etrie
Coh\'erente pour l'Espace) \cite{Bouyer2009,Geiger2011natcomm} as well as the German DLR
funded projects QUANTUS (Quantengase unter Schwerlosigkeit) and PRIMUS
(Pr\"azisionsinterferometrie unter Schwerelosigkeit). Within QUANTUS
interferometry was already demonstrated with degenerate \textsuperscript{87}Rb atoms under
microgravity in the drop tower at ZARM (Germany) \cite{Muentinga_2013,vanZoest2010} and aims
with the MAIUS experiments at realizing quantum gases interferometry on
sounding rockets starting from 2014.

An advantage of using matter waves is that the center of mass positions of the BECs can
be imaged independently for each wave packet and be brought to
coincide. This assumption in the UFF can never be fully matched using classical
bulk matter. At best the deviation caused by initially different positions can
be simulated. The experiment proposed here monitors the motion of two BEC wave packets with initially superposed centers. It can be interpreted as a test of classical general relativity coupled to a Klein-Gordon field in a
non-relativistic limit or, equivalently, a Schr\"odinger equation with
an external gravitational potential.

The paper is organized as follows: In the next Section, the mission concept 
and the expected performance of the AI are outlined. In Sec.~\ref{sec:system}, 
the principle experimental set up is described. The
requirements necessary to achieve the expected accuracy of $2\cdot 10^{-15}$ for
the E\"otv\"os ratio and the error sources are discussed in Sec.~\ref{sec:budgets}.
The planned payload for the STE-QUEST mission is detailed in Sec.~\ref{sec:payload}.


\section{Objectives, Performance and Operation}\label{sec:OPO}
The objective of the STE-QUEST atom interferometer is to test the UFF using
matter waves to an uncertainty in the E\"otv\"os parameter better than $2\cdot
10^{-15}$~\cite{Cacciapouti13SciRD}.
\begin{figure}[tp] \centering
\includegraphics*[width=0.6\textwidth]{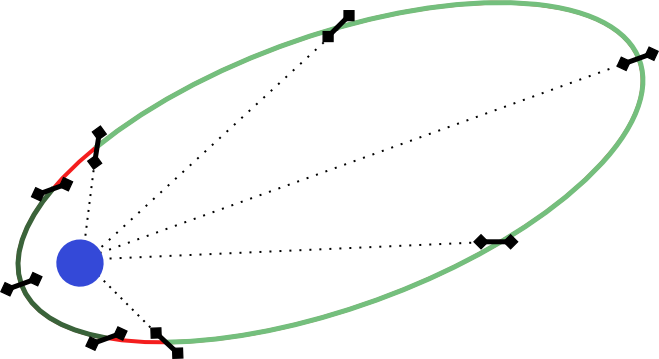} \caption{Highly
elliptical orbit chosen for STE-QUEST clock
comparisons~\cite{Cacciapouti13SciRD}. During perigee pass (dark green) of the
$16\,\mathrm{h}$ per revolution, the spacecraft will be inertially pointing for
$0.5\,\mathrm{h}$ allowing for testing the UFF with the AI part of the payload.
After the perigee phase, the spacecraft orientation is changed to nadir pointing
(red) for clock measurements during the apogee phase (light green). In parallel,
AI parameters are verified and calibrated. The orientation of the atom
interferometer sensitive axis is also indicated in black.}
\label{fig:orbit}
\end{figure}
For STE-QUEST, $\Delta a=a_{87}-a_{85}$ denotes the differential acceleration
between the two wave packets, the sensitivity axis $\vec{e}_{\Delta a}$ is given
by the effective wave vector of the beam splitting light fields
$\vec{k}\parallel\vec{e}_{\Delta a}$. A high common mode rejection ratio for the
differential acceleration of $\approx2.5\cdot10^{-9}$ is a driving requirement
for the overall performance. This and the heritage from various precision and
mixture experiments motivated the choice of \textsuperscript{87}Rb and
\textsuperscript{85}Rb as atomic species for STE-QUEST.
Following~\cite{Damour_2012}, which is one candidate theory describing
violations of the UFF, one would expect an approximately 10--30 times larger
violation of the UFF for other choices of isotopes like \textsuperscript{87}Rb
and K.
However, for these the common mode rejection rejection ratio would be
$\approx300$ for a vibrational background acceleration comparable to
STE-QUEST~\cite{Bouyer2009}.
Thus, although the violation might be smaller the better common mode rejection
for the choice of \textsuperscript{85/87}Rb counter-balances this effectively
turning it into the superior choice.
Compared to state of the art torsion balance~\cite{Schlamminger08PRL} and LLR
tests~\cite{Williams2012CQG} as well as planned or proposed satellite
missions~\cite{Touboul12CQG,Overduin12CQG,Nobili12CQG} with macroscopic test
masses, STE-QUEST offers a complementary approach as a test with a quantum
sensor. Several advantages over proposed ground based atom interferometer
experiments~\cite{Kasevich2007,Dickerson2013PRL,Hogan2009proc} are present due
to the "free fall" conditions in a space borne apparatus. Here, the center of
mass of the atoms is at rest with respect to the experimental set-up.
Consequently, long free evolution times $2T=10\,\mathrm{s}$ can be realised in a
compact set-up. This is a key ingredient to reach a high sensitivity to
accelerations $\vec{a}$, because the phase shift in the interferometer scales as
$\phi_{acc}=\vec{k}\cdot\vec{a}T^{2}$ with the wave number $k$. For ground based
experiments, suspension techniques~\cite{Hamilton2012APS} or large momentum beam
splitters~\cite{Kasevich2007,Dickerson2013PRL,Hogan2009proc} are proposed to
reach high scaling factors although additional constraints due to systematic
errors have to be expected~\cite{Clade2010EPJD}. Using a satellite with inertial pointing mode avoids the necessity of a mirror counter rotation to maintain the interferometer contrast~\cite{Dickerson2013PRL,Lan2012PRL}. Residual rotations of the satellite~\cite{Cacciapouti13SciRD} are compatible with the requirements of the STE-QUEST atom interferometer. Moreover, the low background
accelerations of $4\cdot10^{-7}\,\mathrm{m\,s}^{-2}$ in STE-QUEST compared to
$9.8\,\mathrm{m\,s}^{-2}$ on ground reduce systematic effects and enable the use
of weak traps during the preparation of the atomic ensembles. This is mandatory
to reach atom numbers of $10^{6}$ in dilute ensembles and to efficiently apply
delta kick cooling techniques~\cite{Chu1986,Ammann1997,Morinaga1999,Muentinga_2013} to reach low expansion
rates. Furthermore, a symmetric beam splitting technique~\cite{Leveque2009PRL,Dubetsky2002PRA}
can be implemented which inherently suppresses systematic errors and associated
noise sources. An additional distinctive advantage is the satellite motion which causes a modulation of a possible violation signal. As a result reorientation and rescaling of the gravity signal allow a separation of a possible violation signature from systematic biases that can always occur in a static terrestrial instrument. Systematic errors which are stable in time and do not depend on the
Earth's gravity field can thus be estimated and ruled out.

In STE-QUEST, a quantum projection noise limited sensitivity per cycle of
$\sigma_{\Delta a}/\sqrt{T_{c}}\approx3\cdot10^{-12}\,\mathrm{m\,s}^{-2}$ for $10^{6}$ atoms
of each species, a wave number $k=8\pi/(780\,\mathrm{nm})$, a free evolution
time $T=5\,\mathrm{s}$, and a cycle time $T_{c}=20\,\mathrm{s}$ is anticipated. This value assumes a contrast $C=0.6$. It
is affected by a dephasing due to Earth's gravity gradient $T_{gg}$ coupled to
the initial size $w_r=300\,\mu\mathrm{m}$ and expansion rate $w_v=82\,\mu\mathrm{m\,s}^{-1}$ of the atomic
ensembles and is estimated by the formula $C=\exp\{-\left(kw_{r}T_{gg}T^2
\right)^{2}/2 \}\cdot\exp\{-\left(kw_{v}T_{gg}T^3 \right)^{2}/2
\}$~\cite{Tackmann2012NJP}.

The STE-QUEST AI will measure for $0.5\,\mathrm{h}$ during each perigee pass of
the highly elliptical orbit with a total duration of $16\,\mathrm{h}$  (see
Fig.~\ref{fig:orbit}). At perigee, the proximity to Earth maximizes the signal
of an eventual UFF-violating signal. The satellite will be non-rotating during
this phase which leads to a varying projection of the local gravitational
acceleration $g$ and of the gravity gradient $T_{gg}$ onto the sensitive axis.
Additionally, the interferometer contrast increases as the projection decreases.
The altitude at perigee increases periodically during the mission from about
$700\,\mathrm{km}$ to $2200\,\mathrm{km}$ and then decreases back to
$700\,\mathrm{km}$. An integrated sensitivity per revolution to the E\"otv\"os
ratio of $\sigma_{\eta}^{1\,\mathrm{rev}}\approx5-5.2\cdot10^{-14}$ is expected
when taking into account the shot noise limit, altitude, and attitude of the
satellite with respect to Earth.

Therefore, an integration time of about $1.5\,\mathrm{years}$ is required
to reach the target sensitivity of
$\sigma_{\eta}^{625\,\mathrm{revs}}\approx2\cdot10^{-15}$ compatible with a
total mission duration of $5\,\mathrm{years}$. Residual accelerations of the satellite will be controlled to avoid a signal drift~\cite{Cacciapouti13SciRD}. Parameters of the atom interferometer payload will be re-calibrated or verified during the apogee phase of each orbit to ensure the reproducibility of the UFF test measurements at perigee. Byproduct of the mission will be the most
extended evolution time of cold atoms in a free fall experiment.


\section{System}\label{sec:system}

\subsection{Atom Source}
In order to reach the target performance, a Bose-Bose mixture of $10^{6}$ atoms of each of the isotopes must be prepared in $10$\,s maximum. To this end, an atom chip \cite{Haensel2001,Folman2002,Fortagh2007} setup is used allowing for a fast evaporation and a low power consumption necessary for a satellite-borne device. Moreover, we opt for the use of quantum degenerate ensembles for several reasons. The most important are (i) keeping a reasonably small size of the mixture after a free evolution time of $10$\,s, (ii) reducing the size-related-systematics to an acceptable level and (iii) profiting from the additional control offered by a tunable interactions input state of the atom interferometer. It is important to notice that the dephasing associated to mean-field effects in atom interferometers with interacting sources is reduced here by letting the atomic clouds freely expand until they reach the linear regime of interactions \cite{Posso2013}. Only at this point, the interferometry sequence is started.          

\begin{figure}[tp]
\centering
\includegraphics*[width=1\textwidth]{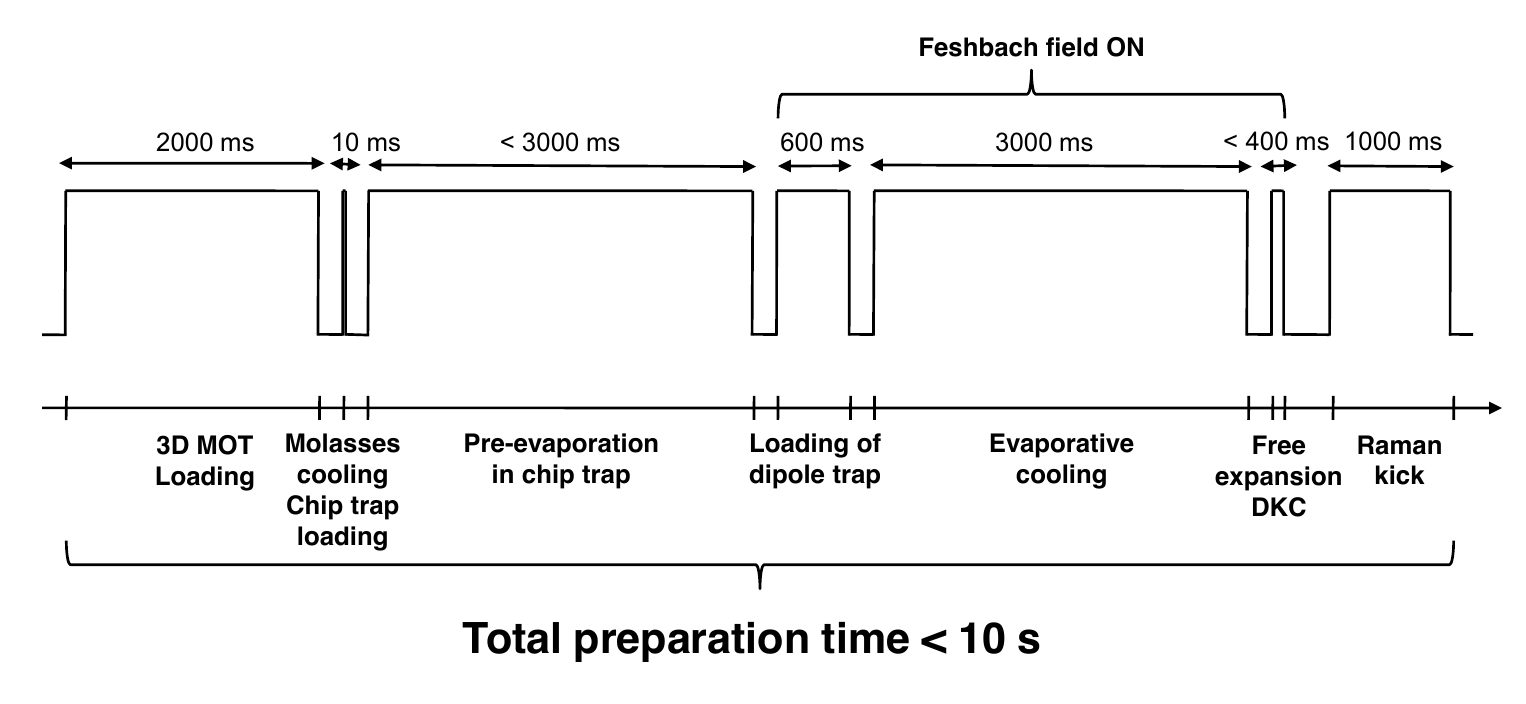}
\caption{BEC generation and preparation sequence. The cycle starts with a loading phase of the main chamber by 
a 2D$^{+}$-MOT. In a rather short time the atoms are trapped in the chip magnetic trap. This allows to pre-evaporate 
the dual source for $3$\,s using RF fields. When the intra-species attractive collisions in $^{85}$Rb start to be
severe due to the increased density of the gas, the Feshbach magnetic field is ramped up tuning these collisions 
to repulsive. The crossed-beam dipole trap is then loaded and the evaporation is performed by lowering the 
power of the tow laser beams. When the two gases reach degeneracy, they are released and freely expand. As soon as the linear regime of interactions is reached, a delta-kick cooling pulse is applied and the Feshbach field is switched-off. The two 
mixed clouds are pushed away from the chip surface by applying a Raman pulse normal to it. The same Raman beam 
is used to stop the atomic clouds when they reach a distance of about $15$\,mm from the surface avoiding the diffraction 
of the interferometry beams on it. This preparation phase is lasting less than
$10$\,s in agreement with the science objectives.}
\label{fig:BECsequence}
\end{figure}

\begin{figure}[tp]
\centering
\includegraphics*[width=1\textwidth]{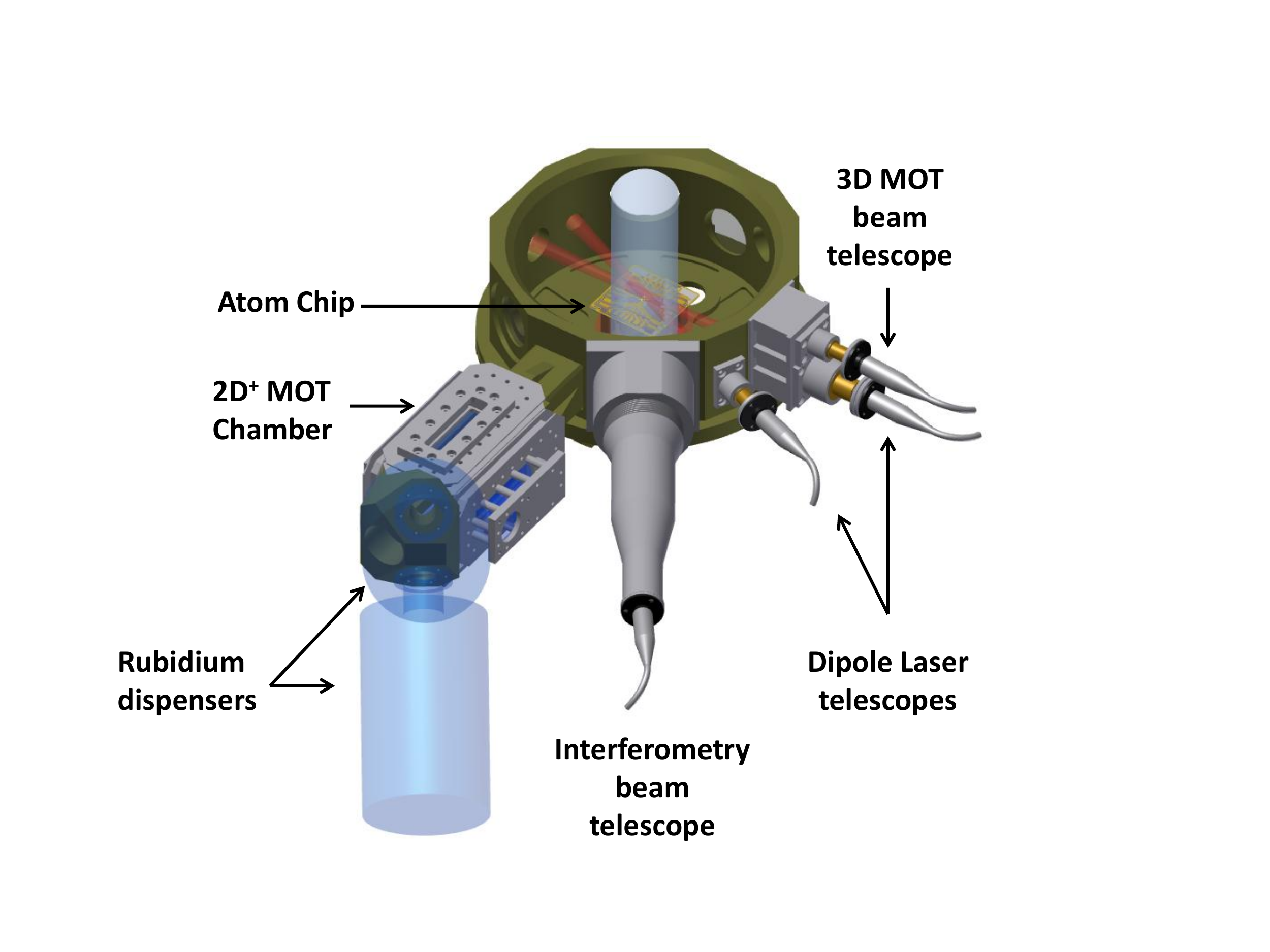}
\caption{Vacuum chambers and main laser beams. A beam of pre-cooled atoms, initially released from the two isotopes reservoirs (light blue), is pushed from the 2D$^{+}$-MOT to the main science chamber via a differential pumping stage. A MOT is loaded right above the center of the atom chip and the atoms are pre-evaporated, after being loaded in the chip trap, using the chip RF antenna. The dipole lasers are depicted by the crossed red beams which trap the atoms at the chip vicinity to finalize the evaporation process. Once the two BECs are obtained, the interferometry pulses are applied by a couple of Raman beams (large light blue beam) along the sensitive axis.}
\label{fig:Chambers_chip}
\end{figure}

The source generation sequence depicted in Fig.~\ref{fig:BECsequence} is
initiated by loading an ultra-high vacuum (UHV) 3D-MOT from a high vacuum (HV) 2D$^{+}$-MOT through a differential pumping stage \cite{Dieckmann1998,Rudolph2011MST} as illustrated in Fig.~\ref{fig:Chambers_chip}. The HV environment is intended for the atomic source, which operates at a Rubidium vapor pressure of a few $10^{-7}$\,mbar. This is the optimal vapor pressure range for the 2D$^{+}$-MOT that provides a pre-cooled beam of atoms towards the UHV chamber. Since the 2D$^{+}$-MOT gains an additional cooling mechanism through the means of two unbalanced counter propagating laser beams along the atom's trajectory, the velocity and the velocity spread of the atoms can be controlled and fast loading ($2$\,s at a flux of $10^{10}$ $^{87}$Rb atoms per second) into the 3D-MOT can be achieved. Thanks to the natural abundance of the $^{85}$Rb isotope ($\geq 72\%$) and the fact that two to three orders of magnitude less $^{85}$Rb atoms (compared to $^{87}$Rb) are necessary at the MOT stage, the same atom source will also be able to generate the envisioned flux of $10^{9}$ $^{85}$Rb atoms per second.

A multi-layer atom chip setup (Fig.~\ref{fig:Chambers_chip}) is used to trap the atoms from the atomic beam. In cooperation with external magnetic bias fields, the chip structures can generate a variety of trap configurations -- from very shallow traps to collect the initial MOT to very tight confinement for fast evaporation \cite{vanZoest2010}. Initially, the largest available (mesoscopic) chip structures are used for the 3D-MOT. For each species, three pairs of counter propagating laser beams, intersecting at the field minimum, are used to generate a mirror MOT \cite{Reichel1999}. The beams contain cooling and repumping light for $^{87/85}$Rb as well. Accordingly, more than $10^{10}$ $^{87}$Rb atoms and $10^{9}$ 
$^{85}$Rb atoms can be captured in a total loading time of 2 seconds.

Once the atoms are captured in the chip MOT, the magnetic fields are
switched-off for a few milliseconds ($5$\,ms)to further cool the atoms through
polarization gradient cooling. The final temperatures of the clouds after all
laser cooling steps will be as low as $20 \mu$K. After switching-on the offset
magnetic field ($5$\,ms), the $^{85}$Rb and the $^{87}$Rb atoms can be
optically pumped to the weak-field seeking states $|F=2,m_{F}=2>_{87}$ and $|F=3,m_{F}=3>_{85}$ in a  fraction of a millisecond.

After state preparation, the lasers are switched-off and the atoms are trapped solely by magnetic fields in a Ioffe-Pritchard trap created by the chip. One exquisite feature of this technology is the ability to generate quite shallow traps (geometric mean of about $7$\, Hz) being in the same time rather deep (around $100\, \mu$K trap depth in the 3 space directions). In this fashion, the atom loss during the MOT trapping is negligible. The temperatures, however, will rise because of heating and adiabatic compression of the trap.

A pre-cooling step is necessary to gain a sufficiently large phase space density
(PSD of $10^{-5}$--$10^{-4}$) before starting the all-optical evaporation. Radio
frequency (RF) radiations are used to pre-evaporate $^{87}$Rb atoms solely. The number of $^{85}$Rb atoms remains approximately constant during this pre-cooling step thanks to the isotope selectivity of these radiations. The $^{85}$Rb atoms cool down sympathetically through collisions with $^{87}$Rb and rethermalize constantly. In about $3$\,s a temperature of a few $\mu$K and a size of about $10\, \mu$m are reached allowing to match the tight confinement of the optical trap and ensure efficient transfer. While the PSD is increased by an order of magnitude, the temperature rises due to an increase in inelastic collisions, especially for $^{85}$Rb atoms. This leads to a loss of one order of magnitude in atom numbers leaving the two ensemble with $10^{9}$ atoms for $^{87}$Rb and $10^{8}$ for $^{85}$Rb left at this step. No further cooling is possible since the 3-body losses of $^{85}$Rb due to its negative scattering length start to be severe at high densities.

Loading the optical trap is costing only another order of magnitude in particles number thanks to the size-compressed and pre-cooled samples. This loading is performed after ramping up a Feshbach field of about $158$\,G in $300$\,ms to avoid disturbing and heating the atoms with eddy currents. This field drives the $^{85}$Rb atoms to a region of positive scattering lengths (ranging from $500\, a_{0}$ to $900\, a_{0}$) to allow for an efficient evaporation \cite{Wieman2008,Altin2010}. Moreover, the magnetic field can be used to change the ratio between elastic and inelastic collisions in $^{85}$Rb and thereby minimize losses by two- and three-body collisions. For all the range of values of the scattering lengths of $^{85}$Rb mentioned above, the two degenerate gases should be in a miscible phase \cite{Wieman2008}. The two ensembles are loaded in a first dipole beam in $300$\,ms followed by a second one with a switch-on duration comparable to the first. Once in place, the final evaporation is carried out. The phase transition to Bose-Einstein condensation (BEC) can be reached in $2-3$\,s using runaway all-optical evaporation \cite{Clement2009}. 
When $10^{6}$ atoms are obtained in the condensed phase for each isotope the far-off resonance lasers are turned-off in $50$\,ms.

An optimization step is starting at this point  and lasts for less than $400$\,ms alternating free expansion and delta-kick cooling (DKC) pulse(s)\cite{Posso2013}. A free expansion of the atomic clouds is starting in the Feshbach field. This expansion phase serves to damp down the density of $^{85}$Rb to a level where the ensemble is stable even in the absence of an external magnetic field \cite{Posso2013}. Not more than a few ms $(3-6)$ are needed to this end. Nevertheless, the bias field is kept for about $10$\,ms after condensation in order to allow the two ensembles to reach the linear regime of interactions and avoid mean-field effects during interferometry. A DKC brief pulse(s) (a fraction of a ms) absorb(s) most of the kinetic energy of the atoms \cite{Chu1986,Ammann1997,Morinaga1999,Muentinga_2013}. This is achieved by suddenly turning -on and -off the final crossed laser traps acting as an atomic lens collimating the BEC clouds to a temperature equivalent expansion of $70$\,pK. This very low effective temperature accessible with DKC is necessary for keeping the size-related systematics at a low level enabling a direct read-out. Fringe patterns building up during the $10$\,s contribute solely to a loss of contrast. An alternative to this low expansion rate is to recover the contrast by unbalancing the time intervals between the interferometry pulses in a suitable way \cite{AlbertContrastPaper}.   
 
Since the last value of the magnetic field tunes solely the scattering length of
$^{85}$Rb, it is possible to optimize its magnitude to reject common size-related error sources such as wave-front curvatures \cite{Posso2013}. This reduces the need for interferometry mirrors from extremely good quality ($\lambda/300$) to values of about $\lambda/30$. At this point the Feshbach field is switched-off without any influence on the free expansion of $^{85}$Rb which recovers its negative scattering length of $-443\, a_0$.

A last manipulation before the interferometry first pulse consists in driving a Raman transition for the atoms in each cloud. One beam normal to the chip and its reflection from the surface are responsible for the 2-photon transition. As a result the atoms travel away from the chip surface. This serves to avoid wave front errors due to the diffraction of the interferometry beams on the chip. In a time interval of $1$\,s, the two ensembles are stopped by reversing the beams at a safe distance of about $15$\,mm.   

\subsection{Interferometery Scheme}

\begin{figure}[tp]
\centering
\includegraphics*[width=0.8\textwidth]{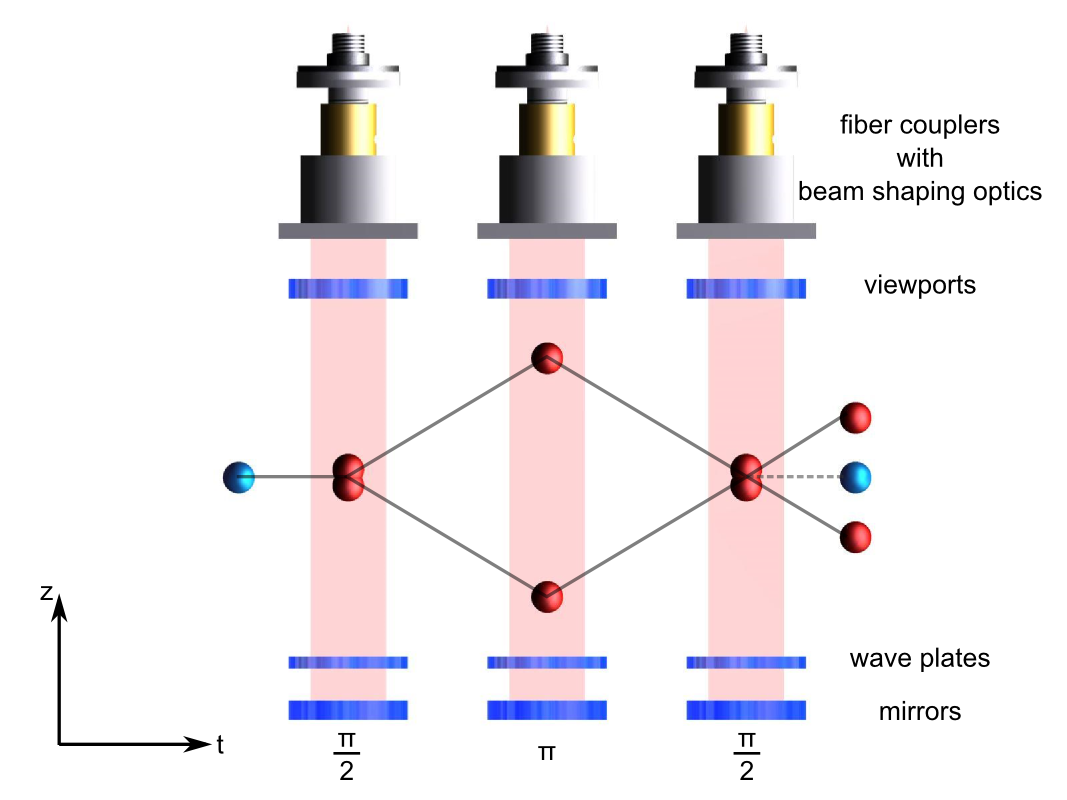}
\caption{Interferometer scheme in a time series. The sequentially applied laser pulses split, reflect, and recombine the atomic wave-functions. The colour of the balls represent the two hyperfine levels of the rubidium atoms. After release, the atoms are in the excited state (red balls), and during the atom interferometry sequence in the ground state (red circles). In this sketch perfect beam-splitting efficiency is assumed.}
\label{fig:int-sequence}
\end{figure}

The interferometer scheme, detailed in~\cite{Schubert13tbs}, is based on a Mach-Zehnder like atom interferometer employing two photon Raman-transitions in a double diffraction setup for the coherent manipulation~\cite{Leveque2009PRL}. The interferometric sequence in this case is composed of a coherent splitting of the wave function into the two interferometer states, a mirroring of theses states after a given interferometer time T and another subsequent splitting after a time T which closes the interferometer and encodes the phase difference between both paths into the population of the output ports. A two photon Raman-transition couples the two hyperfine levels of the rubidium ground state while at the same time transferring a momentum of $2\hbar k_{s}$ to the atoms, where $\hbar k_{s}$ denotes the momentum transfer corresponding to the single photon transition. If the initial state has a vanishing momentum in comparison to the two-photon light field, the two momentum states with $\pm 2\hbar k_{s}$ are degenerated and a splitting into both states will occur as long as the effective momentum transfer is geometrically possible. This is obtained by retro-reflecting the light fields that are driving the Raman-transition. In this scheme, an effective momentum splitting of $4\hbar k_{s}=\hbar k$ is realised while the hyperfine state in the trajectories is always the same. The higher order coupling of the light fields yields to a stronger dependence of the transition probability on the velocity spread of the atomic cloud. Therefore as described in~\cite{Posso2013} an atomic ensemble with an effective temperature of $70\,\mathrm{pK}$ is used as initial interferometer state. Residual occupation of the state $0\hbar k_{s}$ is removed via a resonant light field since the internal state is different to the diffracted orders with $\pm 2\hbar k_{s}$. A sketch of the interferometric sequence can be seen in Fig.~\ref{fig:int-sequence}.
Using a double diffraction scheme reduces the impact of phase shifts dependent on the hyperfine state. Examples are magnetic fields and off-resonant light fields coupling into the interferometer. Magnetic field gradients can still give rise to a residual phase shift. To circumvent this effect, the input hyperfine state can be switched between two successive measurements leading to a reversal of the effective coupling to magnet fields and thus suppressing gradient dependent phase shifts.\\
Gravimetric measurements based on atom interferometers are usually limited by environmental noise, mainly vibrations of the experimental platform. The impact of these accelerations on the interferometer phase is determined by the sensitivity function which is dependent on the effective wave vector $k$, the pulse timings and the Rabi-frequency of the two photon transition \cite{Cheinet2008IEEE}.
As long as these values are matched, environmental noise would lead to the same phase shifts for both species and thus vanish in the differential signal. The interferometer time T and beam splitter pulse duration is set to be equal due to the use of common switching elements for all beams. To match the effective wave vectors and the Rabi-frequencies, the detuning of the Raman-beams to the single photon excitation and the power of the individual beams can be adjusted. The quality of this match directly influences the possible suppression of common mode accelerations (see Sec.~\ref{sec:budgets}) and is discussed in more detail in~\cite{Schubert13tbs}.

\section{Error budget}
\label{sec:budgets}
The choice of \textsuperscript{87}Rb and \textsuperscript{85}Rb is specifically attributed to the engineering of a large common mode rejection ratio. Still, several effects acting differently on the two isotopes can lead to a differential acceleration signal masking a possible violation signal. Additionally, every random fluctuation of a bias term has to stay below shot noise to not impede the targeted uncertainty. A detailed discussion can be found in~\cite{Schubert13tbs}. 
 \begin{table}[tb]
 \begin{center}
 \begin{tabular}{p{0.3\textwidth}p{0.45\textwidth}c}\hline
   Noise source & Conditions & Limit \\ 
  &   &  ($10^{-12}\,\mathrm{m\,s}^{-2}$) \\ \hline
       Shot noise & $10^{6}$ atoms, $C=0.6$ & $2.93$ \\ 
  Linear vibrations & Suppression ratio $2.5\cdot10^{-9}$ & $\approx 1$  \\ 
  Beam splitter laser & Linewidth $100\,\mathrm{kHz}$ & $0.8$ \\
  linewidth & & \\  
  Magnetic fields & $B_{0}=1\,\mathrm{mG}$, $\nabla B_{0}=83\,\mu\mathrm{G\,m^{-1}}$ & $0.11$  \\ 
   Mean field & Beam splitting accuracy $0.001$, & $0.3$  \\
   & $20\,\%$ fluctuation in $N_{87}/N_{85}$ & \\
  Overlap & $10\,\%$ fluctuation per cycle & $<0.1$  \\ \hline
 \hline
  Sum &  & 3.2  \\ \hline
 \end{tabular}
  \caption{Preliminary assessment of statistical errors for the STE-QUEST AI}
 \label{tab:statistical_errors_table}
 \end{center}
 \end{table}
\paragraph{Shot noise and contrast} Both atomic ensembles will feature $N=10^{6}\,\mathrm{atoms}$. The effective wave vector of $k=8\pi/(780\,\mathrm{nm})$, the free evolution time $T=5\,\mathrm{s}$, and the contrast $C=0.6$ are linked to the shot noise limited sensitivity per cycle $\sigma_{\Delta a}/\sqrt{T_{c}}=\sqrt{2/N}\,(CkT^{2})^{-1}\approx2.93\cdot10^{-12}\,\mathrm{m\,s}^{-2}$ for a cycle time $T_{c}=20\,\mathrm{s}$. Herein, the contrast is limited by velocity dependent phase shift in the interferometer coupled to the velocity distribution of the atomic ensemble~\cite{Lan2012PRL,Tackmann2012NJP}. The dominant contribution is given by Earth's gravity gradient $T_{gg}$. Since the orientation and altitude of the satellite with respect to the Earth changes during perigee pass so does the contrast. Here, $C=0.6$ is the minimum for an altitude of $700\,\mathrm{km}$ above Earth and $\vec{k}\parallel\vec{g}$. The rotation rates of $10^{-6}\,\mathrm{rad/s}$ in all three axes do not significantly affect the contrast in STE-QUEST. Velocity selectivity of the beam splitter neither threatens the contrast.
\paragraph{Spurious accelerations of the spacecraft} Any bias acceleration or vibration is suppressed in the differential signal. Suppression ratios of $140\,\mathrm{dB}$ were demonstrated in single species differential atom interferometers~\cite{McGuirk2002}. This cannot directly be transferred to the dual species case, but the response of an atom interferometer to perturbations is well understood~\cite{Cheinet2008IEEE,Bouyer2009}. Thus, the case of STE-QUEST can be modeled and from matching the wave vectors of the two species to $10^{-9}$ and the Rabi frequencies to $10^{-4}$ a suppression ratio of $2.5\cdot10^{-9}$ can be obtained. This assumes the same switching element for both isotopes inherently matching the pulse duration and free evolution times.
\paragraph{Beam splitter laser linewidth} During the beam splitting process, one of the two light fields driving the Raman transition is reflected at the retro reflection mirror while the other is not. Consequently, a time delay between the arrival of the two phase locked laser beams results. This implies a sensitivity to frequency jitter of the beam splitter lasers during the time delay~\cite{leGouet2007EPJD}. For a Lorentzian linewidth of $100\,\mathrm{kHz}$ integrated over the beam splitter pulse duration ($100\,\mu\mathrm{s}$) the noise contribution per shot is estimated to $8\cdot10^{-13}\,\mathrm{m\,s}^{-2}$, well below the STE-QUEST requirements.
\paragraph{Gravity gradients and rotations, photon recoil} In addition to the leading phase term $\propto kT^{2}$ several other phase terms arise due to the specified spurious rotation rates of $10^{-6}\,\mathrm{rad/s}$ in all three axes~\cite{Cacciapouti13SciRD} and Earth's gravity gradient of $T_{gg}\leq 2.5\cdot10^{-6}\,\mathrm{s}^{-2}$ as derived in~\cite{Hogan2009proc,Bongs2006APB}. Most of these terms vanish due to the common mode suppression ratio, but those proportional to differential position and differential velocity of the atoms remain. The gravity gradient will induce a differential acceleration of $T_{zz}\Delta z$ due to an initial center of mass displacement $\Delta z$ between the ensembles and $TT_{zz}\Delta v_{z}$ due to differential center of mass velocity. Spurious rotations $\Omega_{y}$ and $\Omega_{x}$ coupled to differential center of mass velocities $\delta v_{x}$ and $\delta v_{y}$ will lead to differential accelerations $2\Omega_{y}\Delta v_{x}$ and $2\Omega_{x}\Delta v_{y}$. Consequently, the center of mass overlap at the first beam splitter pulse has to be better than $1.1\,\mathrm{nm}$ and $0.31\,\mathrm{nm\,s}^{-1}$ in all three directions. To verify the requirements on relative positioning and velocity of the atomic samples, several images of the atomic ensembles will be taken during the apogee phase with an alternating time of flight of 1\,s and 10\,s after the Raman kick. Fitting the images will reveal the differential center of mass positions. Averaging over a sufficient number of
cycles will then allow a verification at the required precision. This procedure has the same sensitivity in all three direction. Therefore, the same overlap parameters for all three directions are considered. These requirements imply restrictions on the magnetic field gradients during preparation which have to be below $3\,\mu\mathrm{G\,m}^{-1}$. The differential displacement in the optical trap with a trapping frequency of $42\,\mathrm{Hz}$ stays within the defined limit on relative spatial displacement for the gravity gradient of $T_{gg}=2.2\cdot10^{-6}\,\mathrm{s}^{-2}$, and for rotation rates below $1.4\,\mathrm{mrad/s}$ imposing a Coriolis force coupled with the distance to the center of mass of the satellite defensively assumed to be $2\,\mathrm{m}$, magnetic field gradients below $12\,\mu\mathrm{G\,m}^{-1}$, and bias accelerations below $20\,\mu\mathrm{m\,s}^{-2}$. This is compatible with operation both during inertially and nadir pointing phases. Contributions to the differential acceleration signal due to payload and spacecraft self-gravity will be subtracted by comparing perigee and apogee measurements. In first order, the gravity gradients are dominated by the Earth's contribution.
\paragraph{Magnetic fields} During interferometry, both isotopes are in the
magnetic substate $m_{f}=0$ to exclude a linear Zeeman shift. Still, the
quadratic Zeeman effect coupled to the small offset field $B_{0}=1\,\mathrm{mG}$
and a magnetic field gradient $\nabla B$ induce an
acceleration~\cite{leGouet2008APB,Steck2008_Rb87}. Since the coefficients for
the quadratic Zeeman effect are different for the two isotopes, a differential
acceleration signal results. This also impedes the overlap during the time
between release from the optical trap and the delta kick and requires on
magnetic field gradients below $3\,\mu\mathrm{G\,m}^{-1}$. Efficient suppression of external fields is shown in \cite{Milke_2013}.
\paragraph{Effective wave front curvature} When the atomic ensembles expand
in the time interval between two successive interactions with a curved effective beam splitter wave front a phase shift appears~\cite{Louchet2011NJP}. This effect is suppressed in the differential signal because of the similar expansion rates of the two ensembles. In Table~\ref{tab:error_budget_table} the curvature of the retro reflector is assumed to be $R=250\,\mathrm{km}$ and the resulting effective wave front for an initial collimation of the beam splitter telescope $400\,\mathrm{m}$. By matching the expansion rates, the requirements on $R$ will be reduced to be compatible with a retro reflection mirror surface planarity of $\lambda/50$.
\paragraph{Mean field} Even in the regime of linear expansion there is a residual contribution from the mean field energy. This appears in the interferometer signal if the beam splitting at the first pulse is not perfect~\cite{Debs2011PRA}. To mitigate this effect, the mean ratio between the two isotope populations will be tuned to $N_{87}/N_{85}\approx1.697(\pm 0.001)$. Thus, negative energy shift due to \textsuperscript{85}Rb intra species interactions and positive energy shifts due to inter species and \textsuperscript{87}Rb intra species interactions nearly cancel with a remaining uncertainty of $2\cdot10^{-15}\,\mathrm{m\,s}^{-2}$. It is important to stress, that the ratio is allowed to fluctuate at a level of 20\,\% between measurements but has to be controlled in the average to the given level.  The stability of this ratio is continuously monitored as a byproduct of the detection scheme and can be tuned by changing the loading and evaporation parameters of the source. 
\paragraph{Detection efficiency} Vibrations will wash out the fringe visibility, but the differential signal can still be extracted from an ellipse fitting technique~\cite{Foster2002OL,stockton2007bed}. If the outputs of the two atom interferometers are not balanced by a factor $\epsilon$, this will be misinterpreted as an acceleration signal. The parameter $\epsilon$ can be estimated within parts per thousand contributing an error below $10^{-15}\,\mathrm{m\,s}^{-2}$.
 \begin{table}[tb]
 \begin{tabular}{lll}
 \hline
   Error source & Limit &Conditions  \\
   &  ($10^{-15}\,\mathrm{m\,s}^{-2}$) &\\ \hline
  Gravity gradient$^{1}$
  & $2.6$ & $\Delta z=1.1\cdot10^{-9}\,\mathrm{m}$   \\
  & $3.5$& $\Delta v_{z}=3.1\cdot10^{-10}\,\mathrm{m\,s}^{-1}$   \\
  Coriolis acceleration
  & $0.62$& $\Delta v_{x}=3.1\cdot10^{-10}\,\mathrm{m\,s}^{-1}$\\ 
 & $0.62$
  & $\Delta v_{y}=3.1\cdot10^{-10}\,\mathrm{m\,s}^{-1}$\\ 
  Additional overlap dependent terms     & $0.055$ & $\Delta x=1.1\cdot10^{-9}\,\mathrm{m}$ \\
   & $0.0016$ & $\Delta y=1.1\cdot10^{-9}\,\mathrm{m}$ \\ 
  Others 
  & $0.046$  & \\ 
  Photon recoil 
  & $0.039$   & Earth's 2nd order gravity gradient \\
  Self-gravity$^{2}$ 
  & $1$  &\\
  Static magnetic fields$^{3}$
   & $1$ & $B_{0}=1\,\mathrm{mG}$, $\nabla B_{0}=1\,\mu\mathrm{G\,m}^{-1}$ \\  
  Effective wave front curvature$^{4}$
  & $0.63$ & Mirror curvature \\ 
   & $0.28$ & $R=250\,\mathrm{km}$, initial collimation $\approx400\,\mathrm{m}$\\
   && $T_{at}\approx0.07\,\mathrm{nK}$ \\
  Mean field 
  & $2$ & Beam splitter accuracy $0.1\,\%$\\
  && $N_{87}/N_{85}\approx1.697(\pm 0.001)$ \\   
  Spurious  accelerations
   & $1$ & Suppression ratio $2.5\cdot10^{-9}$,\\
   && spurious acceleration  $4\cdot10^{-7}\,\mathrm{m\,s}^{-2}$\\ 
  Detection efficiency$^{5}$ 
   & $<1$ & $|\epsilon-1|<0.003$\\
  & &  \\ \hline \hline
 Total diff. acceleration &  $7.9$&   \\ \hline
 \end{tabular}
  \caption{Preliminary error budget for the STE-QUEST AI \\ The differential acceleration of $7.9\cdot10^{-15}\,\mathrm{m\,s}^{-2}$ was evaluated at perigee for an altitude of $700\,\mathrm{km}$ implying a gravity gradient of $2.2\cdot10^{-6}\,\mathrm{s}^{-2}$ and a projection of the local gravitational acceleration of $8\,\mathrm{m\,s}^{-2}$. Dividing the differential acceleration by the projection of local gravitational acceleration leads to the E\"otv\"os ratio. Terms dependent on the overlap and effective wave front curvature were treated as correlated within their subset, while other terms are expected to be uncorrelated.  $^{1}$ Connected to magnetic field gradient and distance to the center of mass. $^{2}$ Calibration during apogee. $^{3}$ Relieved by input state reversal. $^{4}$ Relaxed by expansion rate match. $^{5}$ Post correction from Bayesian fit. }
 \label{tab:error_budget_table}
 \end{table}
\paragraph{Result} The estimated statistical errors compatible with a shot noise limited measurement are stated in Table~\ref{tab:statistical_errors_table}. An overview of the bias errors assessed at perigee for an altitude of $700\,\mathrm{km}$ is given in Table~\ref{tab:error_budget_table}. Herein, the differential acceleration of $7.9\cdot10^{-15}\,\mathrm{m\,s}^{-2}$ has to be divided by the projection of local $g\approx8\,\mathrm{m\,s}^2$ which leads to an error in the E\"otv\"os ratio of $1\cdot10^{-15}$. During the arc at perigee, the projection of the Earth's gravity gradient and local gravitational acceleration change implying an increase in the uncertainty to $2\cdot10^{-15}$ at the edges. The maximum perigee altitude of $2200\,\mathrm{km}$ and the corresponding arc inhibit the same uncertainty figures.\\
A crucial point to stay within error budget is the initial overlap and differential velocity which will be measured by spatial imaging during each orbit around apogee. The specified gravity gradient, rotation rates, and magnetic field gradients which could cause a displacement in the optical trap combined with a distance to the satellite's center of mass below $2\,\mathrm{m}$ are compatible with the performance budget presented in Table~\ref{tab:error_budget_table}.

\section{Payload}\label{sec:payload}

\begin{figure}[tp]
\centering
\includegraphics*[width=0.8\textwidth]{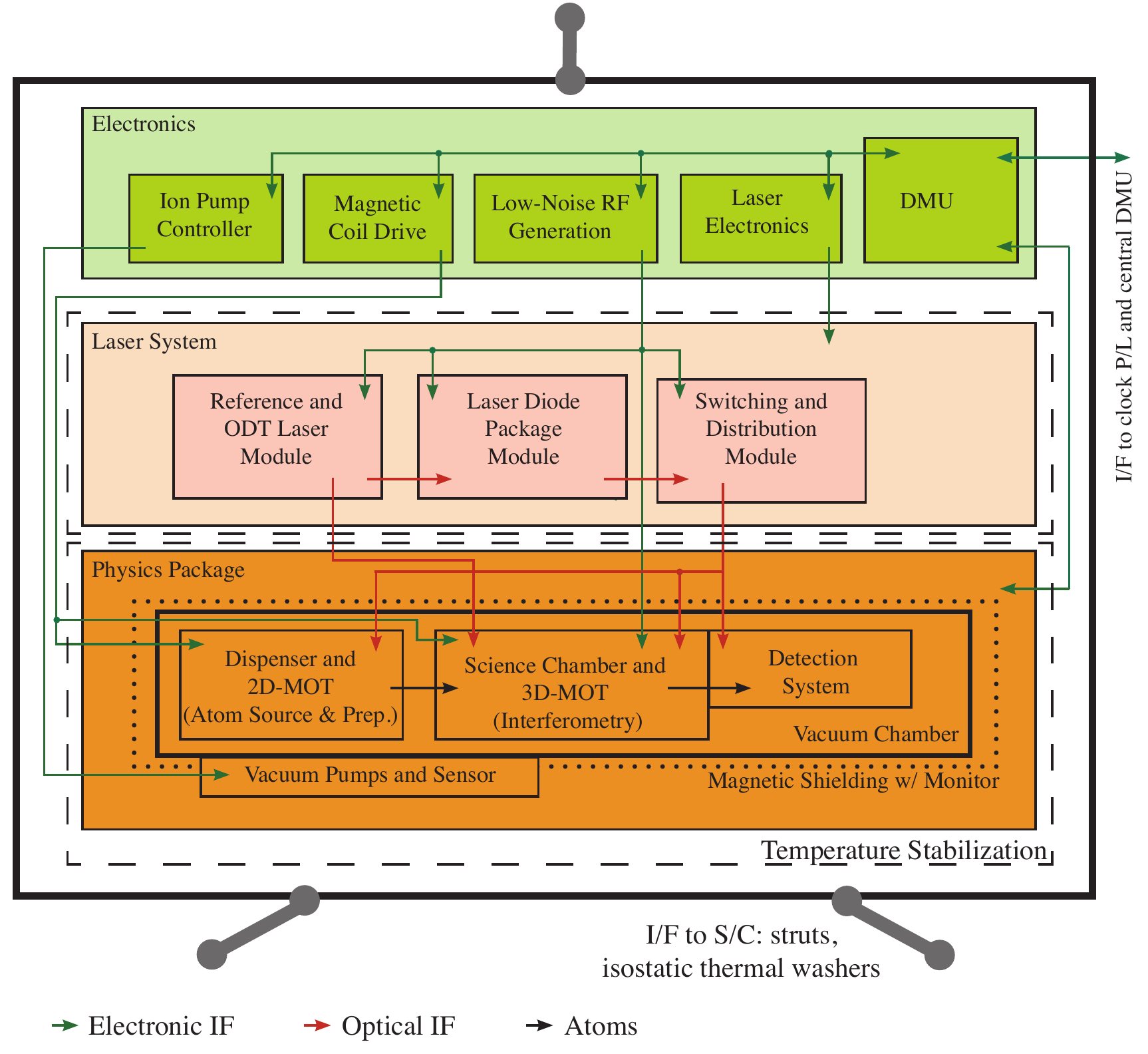}
\caption{Functional diagram of the STE-QUEST atom interferometer payload. It consists of physics package (PP), laser system (LS) and electronics with given subsystems and interfaces.}
\label{fig:functionaldiagram}
\end{figure}

The STE-QUEST atom interferometer payload is subdivided into three main functional units: (i) physics package (PP), laser system (LS) and (iii) electronics as shown in the functional diagram given in Figure\,\ref{fig:functionaldiagram}. The overall preliminary budgets concerning volume, mass and power are detailed in Table\,\ref{table:budgets}. Furthermore, a telemetry budget of 110\,kbps is allocated to the AI. The instrument design is based on current state-of-the-art cold atom experiments under microgravity, namely the German funded QUANTUS (QUANTengase Unter Schwerelosigkeit) and MAIUS (MAteriewelleninterferometrie Unter Schwerelosigkeit) projects operated in drop tower experiments and the French funded I.C.E. (Int\'erf\'erom\'etrie Coh\'ehente pour l'Espace) project operated in zero-g parabola flights. 

\begin{table}[h!]
\centering
\begin{tabular}{lccccc}
\hline
& Volumes& Volume & Mass & Average power & Peak power\\
& & (l) & (kg) & (W) & (W)\\
\hline
Physics Package &1 cylinder & 342 & 135 & 74 & 157\\
Laser System & 3 boxes & 59 & 52 & 103 & 114\\
Electronics & 5 boxes & 68 & 34 &  431 & 549\\
\hline
Total  && & 221 & 608 & 820\\
\hline
\end{tabular}
\caption{Preliminary budget table of the STE-QUEST atom interferometer payload detailing volume, mass and power for the three functional units. All numbers for mass and power include a 20\% component level margin but no system level margin.}
\label{table:budgets}
\end{table}

The Physics Package comprises the Titanium made vacuum chamber for cold atom
preparation and manipulation including atom source, ultra-high vacuum science
chamber, detection unit, vacuum pump system and Mu-metal magnetic shielding.
The science chamber houses the three layer atom chip and features a dodecagon
design providing the optical accesses for optical dipole trap (ODT), 3D-MOT,
interferometry, fluorescence and absorption detection and Raman kick beams. The
atom source consists of a heated Rb reservoir and a 2D-MOT which is attached to
the science chamber using diffusion brazing. The homogeneous magnetic offset
fields are generated using three pairs of coils in Helmholtz configuration. A
four layer Mu-metal shielding with a suppression factor $>10.000$ is foreseen
around the physics package in order to suppress external magnetic stray fields.
The shielding also has to withstand magnetic fields up to 160\,G (Feshbach
field) without permanent damage. The pump system needs to maintain an ultra-high
vacuum at the 10$^{-11}$\,mbar level and uses a combination of an ion getter
pump and a passive getter pump.

The Laser System is housed in three boxes: (i) a Telecom fiber technology based
reference and optical dipole trap laser module, (ii) a micro-integrated,
high-power 780\,nm laser diode package module for atom manipulation, cooling and
detection and (iii) a switching and distribution module delivering the laser
beams according to the experimental sequence (cf.~Sec. 3) to the physics
package. The switching module is based on Zerodur bonding technology and uses a
combination of acousto-optic modulators (AOMs) for fast switching and mechanical
shutters for highest extinction ratio, while the distribution module is realized
as an optical fiber technology beam splitter array.

The AI instrument electronics includes a data management unit (DMU) which
controls all other electronics units and the overall payload, including
housekeeping data gathering, a magnetic coil drive unit providing the low noise
current drivers for magnetic field generation, a low-noise RF generator
producing the 6.8\,GHz and 3\,GHz signals corresponding to the hyperfine
transitions in \textsuperscript{87}Rb and \textsuperscript{85}Rb, respectively
and the signals for RF knife and driving electro-optical components, a laser
control unit  providing the low noise current supplies and temperature controls
for the lasers, and an ion pump controller delivering the high voltage power
supply for the ion getter pump.

\section{Conclusion}
The STE-QUEST mission aims to perform a quantum test of the
Universality of Free Fall using cold atom interferometry with unprecedented precision, exploring in this way the frontiers of the validity of General Relativity. 
The mission will track the propagation of two matter waves of atomic species, i.e. two Bose Einstein condensates consisting of \textsuperscript{85}Rb and \textsuperscript{87}Rb, which
fall freely in Earth's gravitational field.
The goal of the mission is to reach an accuracy of  $\eta\leq 2\cdot 10^{-15}$ over the entire mission
period, improving the best test performed on Earth so far by at least two orders of magnitude. 
With this accuracy a new window is opened to find experimental evidence of a
quantum theory of gravity -- today's main open question in theoretical physics.

\ack
This work was supported by the German space agency ``Deutsches Zentrum f\"ur
Luft- und Raumfahrt (DLR)'' with funds provided by the Federal Ministry of
Economics and Technology under grant numbers 50 OY 1302, 50~OY~1303, and 50 OY
1304, the German Research Foundation (DFG) by funding the Cluster of Excellence
``Centre for Quantum Engineering and Space-Time Research (QUEST)'' and the
research traing group ``Models of Gravity'', the French Space Agency Centre
National d'Etudes Spatiales, and the European Space Agency (ESA).

\section*{References}
\bibliographystyle{h-physrev}
\bibliography{STE_QUEST_MissionPaper}

\newpage


\end{document}